\documentclass[11pt]{article}
\usepackage{epsfig,psfig,float,amssymb,latexsym}
\usepackage[notref,notcite]{}
\newcommand{\NP}[1]{ Nucl.\ Phys.\ {\bf #1}}
\newcommand{\ZP}[1]{ Z.\ Phys.\ {\bf #1}}

\newcommand{\PL}[1]{ Phys.\ Lett.\ {\bf #1}}

\newcommand{\PR}[1]{Phys.\ Rev.\ {\bf #1}}
\newcommand{\PRL}[1]{ Phys.\ Rev.\ Lett.\ {\bf #1}}

\newcommand{\bi}{\bibitem}
\newcommand{\vs}{\vspace{-0.20cm}}
\newcommand{\Opc}{\mathcal{O}(p^4)}

\newcommand{\be}{\begin{equation}}
\newcommand{\ee}{\end{equation}}
\newcommand{\ba}{\begin{eqnarray}}
\newcommand{\ea}{\end{eqnarray}}

\newcommand{\nn}{\nonumber}

\begin{document}
%
%
\begin{center}
{\Huge{\bf{ Pion and Kaon Vector Form\\
\vspace{0.4cm}
 Factors}}}
\end{center}
\vspace{.5cm}

\begin{center}
{\huge{J. A. Oller}}
\end{center}

\begin{center}
{\small{\it Forschungszentrum J\"ulich, Institut f\"ur Kernphysik (Theorie) \\ 
D-52425 J\"ulich, Germany}}
\end{center}

\begin{center}
{\huge{E. Oset and J. E. Palomar}}
\end{center}

\begin{center}
{\small{\it Departamento de F\'{\i}sica Te\'orica e IFIC, Centro Mixto
Universidad de Valencia-CSIC,\\
46100 Burjassot (Valencia), Spain}}
\end{center}

\vspace{1cm}

\begin{abstract}
{\small{We develop a unitarity approach to consider the final state interaction 
corrections to the tree level graphs calculated from Chiral Perturbation
Theory ($\chi PT$) allowing the inclusion of explicit resonance fields. The method is
discussed considering the coupled channel pion and kaon vector form factors.
These form factors are then matched with the one loop $\chi PT$ results. A very 
good description of experimental data is accomplished for the vector 
form factors up to $\sqrt{s}\leq
1.2$ GeV beyond which multiparticle states play a non negligible role. For the
P-wave 
$\pi \pi$ phase shifts the agreement with data stands even higher up to
$\sqrt{s}\leq 1.5$ GeV. We also consider the isospin breaking effects due
to the $\omega-\rho$ mixing as a perturbation to the previuos results. In
addition the low and resonance energy regions are discussed in
detail and for the former a comparison with one and two loop $\chi PT$ is made
showing a remarkable coincidence with the two loop $\chi PT$ results.}}  
\end{abstract}

\section{Introduction}
The study of the pion vector form factor is an interesting problem mainly
because pions are the lightest hadrons and hence they are common products in many
experiments so that a good description of the pion electromagnetic
form factor is often required.

        Many of the studies of this problem deal with some kind of modified 
vector meson dominance (VMD) \cite{book} when taking into account the effects of
unitarity and final state interactions. This
was done long time ago in ref. \cite{gou} and it was found that these effects
show up not only 
as a modification of the bare $\rho$ propagator but also as a change in its bare 
couplings. Other more phenomenological parameterizations are the ones in refs. 
\cite{arcadi,barkov}, which basically account for the dressing of the bare $\rho$ 
propagator and allow to add more resonances and parameters in order to fit the data 
up to high energies, hiding in this way extra effects as
the presence of multiparticle channels, e.g. $4 \pi$, $\omega \pi$ etc... This latter 
kind of expressions are the ones commonly used
in many experimental papers in order to fit their data and determine the 
resonance content. 

It is well known that the low energy effective theory of QCD is Chiral
Perturbation Theory ($\chi PT$) \cite{wein,gl1,gl2}. Although this is a
systematic way to express the QCD Green functions in terms of a power
momentum expansion, unfortunately, it is valid only for low energies. Hence, 
if one is interested in higher energies nonperturbative schemes are 
unavoidable. Nevertheless, one should demand that, when used at low energies,
these nonperturbative methods reproduce the low
energy constraints of QCD given by the $\chi PT$ expansion.
 
 The approach described in sections 2 and 3 reproduces the one loop $\chi PT$
 pion and kaon vector form factors \cite{gl2} and, as we 
 show below, it is
 also appropriate to study higher energies by satisfying unitarity in coupled 
 channels and incorporating explicit resonance fields \cite{ecker}. In
 this latter reference it is discussed how, at lowest order in
 the chiral expansion, resonances  with spin $\leq 1$ couple with 
 pseudoscalars ($\pi$, $K$, $\eta$) and with electroweak currents. It is also 
 shown there how to
 connect with the standard VMD picture, when taking into account results from 
large $N_c$ and QCD high energy constraints.

Other nonperturbative approaches that match with the one loop $\chi PT$ vector pion 
form factor 
are given in refs. \cite{toni,hanna,Nieves:2000bx}. In ref. \cite{hanna2,ulf2} the matching is given 
up to the two loop $SU(2)\times SU(2)$ $\chi PT$ pion vector form factor first calculated 
numerically in \cite{ulf2} and then analytically in \cite{cola}. 
While the work of ref. \cite{ulf2} is only interested in assessing the relevance of 
higher order loops at low energies, the works of refs. \cite{toni,hanna,Nieves:2000bx,hanna2} make predictions
at higher energies, although all of them take into
account only elastic unitarity. Ref. \cite{toni}, as ref. \cite{ulf2}, makes use of an 
Omn\`es representation\footnote{Although in the end, since only the real part of the
exponent of the exponential is kept, some of the analytical properties
of the form factor are lost.} \cite{om} but allowing 
for the explicit presence of the $\rho$ resonance \cite{ecker}. In
ref. \cite{Nieves:2000bx} the pion vector form
factor is studied by solving the Bethe-Salpeter equation making some assumptions  on
the off shell extrapolation of the amplitudes and using only the $\pi\pi$
channel. On the other 
hand, in ref. \cite{hanna,hanna2} [0,1] and [0,2] Pad\'e approximants from the $\chi PT$ 
results are used (see also the discussions in refs. \cite{ulf2,guerrista} with respect 
the former references). Finally in refs. \cite{Guerrero:1999ei,Pich:2001pj} an
Omn\`es parameterization, assuming elastic
unitarity, is implemented using directly the experimental I=1 P-wave
$\pi\pi$ phase shifts.

In our work the isospin limit is taken but we shall also estimate the effect of
the isospin violating $\omega-\rho$ mixing in the
pion vector form factor. This 
contribution manifests itself in a very narrow energy range to the
right 
of the peak of the $\rho$ mass distribution in accordance
with the mass and small width of the $\omega$ (see fig.\ref{ffmixing}). Other works 
 in which this effect has been already discussed are  
\cite{barkov,ocko,tesis}. On the other hand, when comparing our results for the vector
pion form factor in the $\rho$ region with the experimental data, we shall also 
 take data from $\tau$-decays \cite{tau} in which 
only the $I=1$ part of the 
vector pion form factor contributes. Finally, in ref. \cite{kubis} the 
isospin breaking 
effects in the pion form factors for the low energy regime are studied at 
next-to-leading order in $SU(2)\times SU(2)$ $\chi PT$.

\section{Tree level}

We will evaluate the tree level vector form factors and scattering amplitudes
making use of the lowest order $\chi PT$ Lagrangian \cite{gl2} plus the chiral
resonance Lagrangians \cite{ecker}. These tree level amplitudes will be leading
in large $N_c$, while non leading contributions in this counting will be generated 
through the 
unitarization process to be discussed in the next section.

\begin{figure}[ht]
\centerline{\includegraphics[width=.9\textwidth]{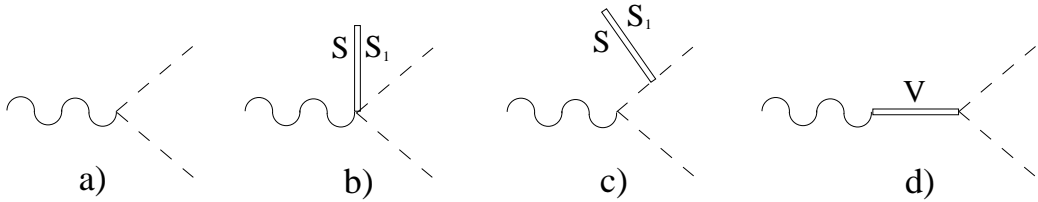}}
\caption{Diagrams considered for calculating the tree level form factors, eqs.
(\ref{fftree}). From left to right: a) $\chi PT$ lowest order. 
b) Octet, $S$, and singlet, 
$S_{1}$, scalar resonance exchanges coupled to the vacuum. c) Wave function
renormalization from the exchange of scalar resonances and d) exchange of 
vector resonances, $\rho$, $\omega$ and $\phi$. Fig. \ref{uno}a is calculated from ref.
\cite{gl2} and the rest from ref. \cite{ecker}.}
\label{uno}
\end{figure}

In the case of the vector $\pi\pi$ and $K\overline{K}$ form factors we have to
evaluate the diagrams depicted in fig.\ref{uno}. The
contributions from the diagram of fig.\ref{uno}b and the ones from the wave function 
renormalization, fig.\ref{uno}c, cancel each other due to charge conservation. A similar set 
of diagrams has been considered in ref. \cite{jamin} in order to study the coupled $K\pi$, 
$K\eta$ and $K\eta'$ scalar form factors \cite{jamin2}. In the former reference, 
although restricted to the study of the associated S-wave I=1/2 meson--meson $T$-matrix, 
the requirement that the form factors vanish at $s\rightarrow \infty$ is used to
reduce the number of free parameters.

On the other hand, we will assume ideal mixing between the $\omega_8$ and $\omega_1$ resonances, 
so that the $\omega$  and $\phi$ are given by
\ba
\label{idealmix}
\omega=\frac{2}{\sqrt{6}} \omega_1+\frac{1}{\sqrt{3}} \omega_8 \nn \\
\phi=\frac{1}{\sqrt{3}} \omega_1-\frac{2}{\sqrt{6}} \omega_8
\ea
and we will work explicitly in terms of the $\omega$ and $\phi$ mesons. In this way we will be able to distinguish between the different physical masses
of the $\omega$ and $\phi$ resonances. Note that at the order considered in ref.
\cite{ecker} vector singlet resonances, $\omega_1$, do not couple neither to 
photons nor to pseudoscalars.

The resulting tree level vector form factors for the $\pi^+\pi^-$, $K^+ K^-$ and 
$K^0 \overline{K}^0$ systems are:

\ba
\label{fftree}
F^t_{\pi\pi}(s)&=& 1+ \frac{F_V \,G_V}{f^2}\frac{s}{M_\rho^2-s}\nn \\
F^t_{K^+ K^-}(s)&=& 1+ \frac{F_V \,G_V}{2\,f^2}\, s \left[\frac{1}{M_\rho^2-s}+
\frac{1}{3}\frac{1}{M_\omega^2-s}+\frac{2}{3}\frac{1}{M_\phi^2-s}\right] 
\nn \\
F^t_{K^0
\overline{K}^0}(s)&=&\frac{F_V\,G_V}{2\,f^2}\, s\left[-\frac{1}{M_\rho^2-s}+
\frac{1}{3}\frac{1}{M_\omega^2-s}+\frac{2}{3}\frac{1}{M_\phi^2-s}\right]
\ea
where $F_V$ measures the strength of the photon-vector resonance vertex, $G_V$ 
the same but for the pseudoscalar-vector resonance ones and $s$ is the usual
Mandelstam variable. Finally, $M_\rho$, $M_\omega$ and $M_\phi$ refer to the
bare masses of the $\rho$, $\omega$ and $\phi$ resonances and $f$ is the pion decay
constant, $f_\pi$, in the chiral limit \cite{gl2}.

\begin{figure}[ht]
\centerline{\includegraphics[width=0.9\textwidth]{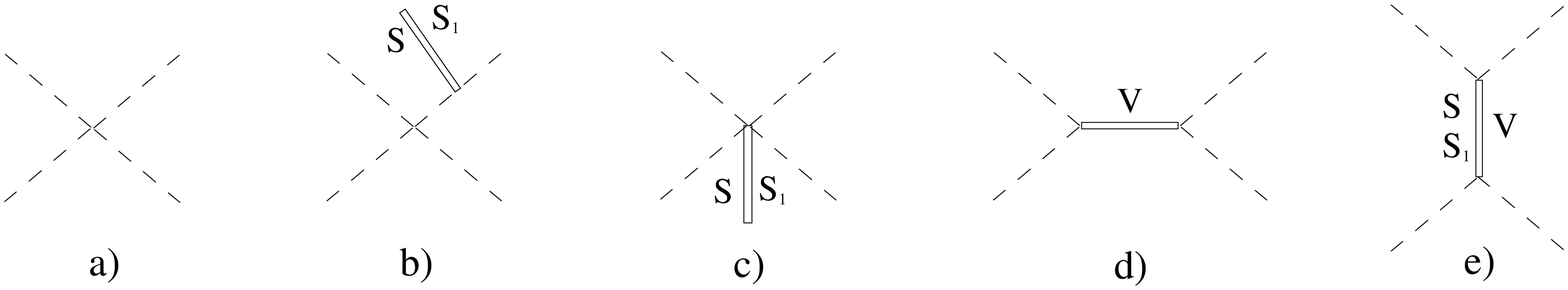}}
\caption{Diagrams considered for calculating the tree level amplitudes, eqs.
(\ref{treeT}). From left to right: a) $\chi PT$ lowest order. 
b) Wave function
renormalization from the exchange of scalar resonances. c) Tadpole-like diagram with the
exchange of scalar resonances coupled to the vacuum. d) Exchange of 
vector resonances, $\rho$, $\omega$ and $\phi$, in the $s$-channel and e)
crossed exchange of vector and scalar resonances. Fig. \ref{dos}a is calculated from 
ref. \cite{gl2} and the rest from ref. \cite{ecker}.}
\label{dos}
\end{figure}

One can proceed analogously for the evaluation of the tree level scattering
amplitudes between the former states. In ref. \cite{bmk} similar tree level amplitudes 
were already calculated in the same way for studying the elastic 
$\pi\pi$ and $K\pi$ scattering. In addition, in ref. \cite{jamin} the $K\eta$ and $K\eta'$ 
channels were also included to study the $K\pi$ scattering up to 2 GeV. The corresponding generic 
set of diagrams is indicated in fig.\ref{dos}. 
However, out of 
these diagrams we are not going to consider explicitly here those corresponding 
to the 
exchange of resonances in the crossed channels, fig.\ref{dos}e. The reason is 
twofold: 1) they
are not necessary to match with the one loop $\chi PT$ results for the $SU(3)$
vector form factors \cite{gl2} since they give rise to higher orders. 2) Because of VMD one 
expects a good
description of the scattering amplitudes in the physical region when including the
$s$-channel exchange of the vector resonances together with unitarity. With
respect to  the second point one has to
take into account that in the chiral Lagrangians \cite{gl2,ecker} the standard picture 
of VMD \cite{sakurai} is only accomplished when considering the lowest order 
$\chi PT$ contribution together with the explicit resonance fields \cite{nd,toni}. 
Furthermore, when making a
dispersive analysis of the vector form factors, only the scattering amplitudes 
in the physical region are involved (see next section).

The resulting P-wave \footnote{A partial wave amplitude with angular
momentum $L$ is defined to be ${\displaystyle{\frac{1}{2} \int_{-1}^1 d \cos
\theta P_L(\cos \theta) T(s,t,u)}}$ where $P_L(\cos \theta)$ is the Legendre
polynomial of $L_{th}$ degree, $T(s,t,u)$ is the scattering amplitude and $s$, $t$ and $u$ are
the usual Mandelstam variables.} 
partial wave amplitudes are:

\ba
\label{treeT}
T(s)_{\pi^+\pi^-,\pi^+\pi^-}&=& \frac{2}{3}\frac{p_\pi^2}{f^2} \left[1+
\frac{2\,G_V^2}{f^2} \frac{s}{M_\rho^2-s}\right]\nn \\
T(s)_{ \pi^+ \pi^-,K^+K^-}&=&\frac{p_\pi \,p_K}{3\,f^2}\left[1+\frac{2\,G_V^2}{f^2}
\frac{s}{M_\rho^2-s}\right] \nn \\
T(s)_{\pi^+\pi^-,K^0 \overline{K}^0}&=&-T(s)_ {\pi^+\pi^-,K^+K^-}\nn \\
T(s)_{K^+ K^-,K^+ K^-}&=&\frac{2}{3}\frac{p_K^2}{f^2} \left[1+
\frac{G_V^2}{2\,f^2} \frac{s}{M_\rho^2-s}+\frac{G_V^2}{2\,f^2}\frac{s}{M_\omega^2-s}+
\frac{G_V^2}{f^2}\frac{s}{M_\phi^2-s}\right] \nn \\
T(s)_{K^0 \overline{K}^0,K^0 \overline{K}^0}&=&T_{K^+ K^-,K^+ K^-}(s)\nn\\
T(s)_{K^+ K^-,K^0 \overline{K}^0}&=& \frac{p_K^2}{3\,f^2}\left[1+
\frac{G_V^2}{f^2}\frac{s}{M_\omega^2-s}+\frac{2\, G_V^2}{f^2}
\frac{s}{M_\phi^2-s}-\frac{G_V^2}{f^2}\frac{s}{M_\rho^2-s}\right]
\ea
By writing the lowest order amplitudes in terms of $f$, there is a 
cancelation between the contributions of the wave function renormalization 
terms, fig. \ref{dos}b, those of the exchange of the scalar resonances coupled to the 
vacuum, fig. \ref{dos}c, and the $\Opc$ crossed channel scalar contributions 
absorbed in the $\Opc$ $\chi PT$ term proportional to the $L_5$ 
counterterm \cite{gl2,ecker}, fig. \ref{dos}e. Consistently with our approach of
working the large $N_c$ leading contributions from ref. \cite{gl2}, we have 
taken the masses of the scalar singlet and octet equal and we have also made 
use of the relation between their couplings deduced in ref. \cite{ecker}. In 
this way, for instance, one has that $L_4=L_6=0$ consistently with 
phenomenology \cite{bijnens} and with the fact that both are subleading counterterms 
in large $N_c$.

\section{Unitarization}
We will deduce our final expressions for the form factors after deriving those of the 
$T$-matrix. The latter will be accomplished by unitarizing the tree level scattering
amplitudes in eqs. (\ref{treeT}) following ref. \cite{nd}. In this
reference  the most general structure of a partial wave
amplitude when the unphysical cuts are neglected was deduced. It was also shown that
this structure is the one required in order to match with the tree level
amplitudes calculated from the lowest order $\chi PT$ Lagrangian \cite{gl2} plus the
exchange of the resonances \cite{ecker}, as in eqs. (\ref{treeT}). In refs. \cite{ww,ulf} the method is
extended to include unphysical cuts as calculated in one loop $\chi PT$ 
with and without baryons, respectively. For a review on these techniques, see ref. 
\cite{review}.

We work in the isospin limit. Our isospin states are:

\ba
\label{isospin}
I&=&1 \nn\\
|\pi\pi>&=&\frac{1}{2}|\pi^+\pi^- - \pi^- \pi^+>\nn \\
|K\overline{K}>&=& \frac{1}{\sqrt{2}} |K^+K^- - K^0\overline{K}^0>\nn \\
&& \nn\\\
I&=&0 \nn\\
|K\overline{K}>&=&\frac{1}{\sqrt{2}} |K^+ K^- + K^0\overline{K}^0>\nn \\
\ea
note the extra factor $1/\sqrt{2}$ in the normalization of the $|\pi\pi>$ $I=1$ vector
since in this state pions behave like identical particles.  We have 
also removed a global $e^{i\pi}$ phase in all the states in order to have the form 
factors positive defined in $s=0$.

        Since for $I=1$ we have two coupled channels, we will use in the following a matrix 
notation in which the pions are labeled by 1 and kaons by 2 in the $I=1$ case. For $I=0$ we 
have only the kaon channel which we denote by 1.

        Taking into account eqs. (\ref{treeT}) and (\ref{isospin}) we can
calculate for both isospin channels the tree level amplitudes between definite $\pi\pi$
and $K\bar{K}$ isospin states. We will collect these amplitudes, for each isospin channel,
in a $K_I$-matrix. Thus, 
from ref. \cite{nd} we have the following expression for the $T_I$-matrix:

\be
\label{T}
T_{I}(s)= \left[I+K_I(s)\cdot g^I(s) \right]^{-1}\cdot K_I(s)
\ee
where $I$ is the unity matrix and $g^I(s)$ is a diagonal matrix \cite{nd} given by the 
loop with two meson propagators. In dimensional regularization one has explicitly:

\be
\label{g(s)}
g^I_i(s)=\frac{1}{16\,\pi^2}\left[-2+d^I_i+\sigma_{i}(s)\,
\log \frac{\sigma_{i}(s)+1}{\sigma_{i}(s)-1} \right]
\ee
where the subindex $i$ refers to the corresponding two meson state and
$\sigma_{i}(s)=\sqrt{1-4 m_i^2/s}$ with $m_i$ the mass of the particles in the
state $i$. For the choice $d^I_i=0$ one has $g^I_i(s)=-\bar{J}(s)_{ii}$ with the 
$\bar{J}(s)_{ii}$ functions defined in ref. \cite{gl2} such that 
$\bar{J}(0)_{ii}=0$. 

We now introduce the diagonal matrix $Q(s)_{ii}=p_i(s) \,\theta(s-4\,m_i^2)$, with 
$p_i(s)=\sqrt{s/4-m_i^2}$ the modulus of the c.m. three momentum of the state $i$ and 
$\theta(x)$ the usual Heaviside function. In terms of this matrix we have

\be
\label{S}
S_I(s)=I+\frac{i}{4\,\pi \sqrt{s}}\, Q(s)^{1/2} \cdot T_I(s) \cdot Q(s)^{1/2}
\ee
with $S_I(s)$ the $S$-matrix fulfilling the requirement $S_I(s)\cdot S_I(s)^\dagger=I$.

The electromagnetic meson form factor, $F_{MM'}(s)$, is introduced as follows. The
transition amplitude from a photon, virtual or real, to a pair of mesons is written as:
\be
\label{defiff}
<\gamma(q)|T|M(p)M'(p')>=e\epsilon_{\mu} (p-p')^{\mu}F_{MM'}(s)
\ee
with $q^2=s$, $e$ the modulus of the charge of an electron and $\epsilon_{\mu}$ the 
photon polarization vector.

Unitarity of the $S$-matrix expressed between the $\gamma$ and the $MM'$ states, up to 
${\mathcal{O}}(e^2)$, leads to the relationship:
\be
\label{optiff}
p_{MM'}(s)\, \hbox{Im}F(s)_{MM'}=\sum_{\alpha}F_{\alpha}(s)\frac{p_\alpha(s)}{8\pi
\sqrt{s}} p_{\alpha}(s) \theta(s-4m_{\alpha}^2)\left( T_{\alpha,MM'}^{L=1}(s)\right)^*
\ee
where$^{\ \ *}$  means complex conjugation and the strong amplitudes are projected in the
P-wave. In the following we will remove the supraindex $L=1$ with the understanding that 
any strong amplitude will be in P-wave.

Dividing the former equation by $p_{MM'}(s)$ and taking the complex conjugation of the term in
the right hand side, since this is a real quantity above the threshold of the system
$|MM'>$ where unitarity applies, one has:

\be
\label{uni1}
\hbox{Im}F_{MM'}(s)=\sum_{\alpha}F^*_{\alpha}(s) 
\frac{p_\alpha (s)}{8\pi\sqrt{s}}\theta(s-4m_{\alpha}^2) p_\alpha (s) 
\frac{T(s)_{\alpha,MM'}}{p_{MM'}(s)}
\ee

Taking into account that the $T(s)_{\alpha \beta}$ is a symmetric matrix with respect  
indexes $\alpha$ and $\beta$ because of time reversal invariance, introducing also the
diagonal matrix $\widetilde{Q}(s)_{ij}=p_i(s) \,\delta_{ij}$ and the one column matrix 
$F(s)_i=F_i(s)$, we can write eq. (\ref{uni1}) in a matrix notation as:

\be
\label{uni2}
\hbox{Im}F(s)=\widetilde{Q}(s)^{-1}\cdot T(s) \cdot \frac{Q(s)}{8 \pi \sqrt{s}} 
\cdot \widetilde{Q}(s)\cdot F^*(s)
\ee

If we further substitute in the previous equation $\hbox{Im}F(s)$ by 
$(F(s)-F^*(s))/(2\, i)$ and $T(s)$ by its expression in eq. (\ref{T}), one finds:

\ba
\label{uni3}
F(s)&=&\left\{I+\widetilde{Q}(s)^{-1}\cdot \left[I+K(s)\cdot g(s)\right]^{-1}\cdot K(s) \cdot
i \frac{Q(s)}{4\pi \sqrt{s}}\cdot
\widetilde{Q}(s) \right\}\cdot F^*(s)\nonumber \\
&=&\widetilde{Q}(s)^{-1}\left[I+K(s)\cdot g(s)\right]^{-1}\cdot \left\{\left[I+
K(s)\cdot g(s)\right]\cdot
\widetilde{Q}(s)+K(s)\cdot i\frac{Q(s)}{4 \pi \sqrt{s}}\cdot \widetilde{Q}(s)\right\}
\cdot F^*(s)
\nonumber\\
&=&\widetilde{Q}(s)^{-1}\cdot \left[I+K(s)\cdot g(s)\right]^{-1}\cdot \widetilde{Q}(s)\cdot \Big\{
I+\widetilde{Q}(s)^{-1}\cdot K(s)\cdot \widetilde{Q}(s) \cdot g(s)  \nonumber \\
&+&\widetilde{Q}(s)^{-1}\cdot K(s) \cdot 
\widetilde{Q}(s) \cdot i \frac{Q(s)}{4 \pi \sqrt{s}}  \Big\}\cdot F^*(s)
\ea
In the last equality we have made use of the fact that the matrices $Q(s)$, 
$\widetilde{Q}(s)$ and $g(s)$ commute since all of them are diagonal.

Let us a first note that the symmetric matrix $K(s)$, eq. (\ref{treeT}), is a 
matrix of functions with a kinematical cut in its $K(s)_{12}$ matrix element 
between the threshold of channels 1 and 2 because $K(s)_{12}$ is proportional to 
$p_1(s)  p_2(s)$. However, the matrix 
$\widetilde{Q}(s)^{-1} \cdot K(s) \cdot \widetilde{Q}(s)$ has no such kinematical cuts 
since the modulus of any of the three-momenta $p_\alpha$ appears always squared. Note
also that, although $\widetilde{Q}(s)^{-1}$ does not exist for $s$ equal to its value 
for any of the thresholds, by continuity, the product $\widetilde{Q}(s)^{-1} \cdot K(s) \cdot
\widetilde{Q}(s)$ is well defined, and hence $\widetilde{Q}(s)^{-1} \cdot K(s)
\cdot \widetilde{Q}(s)$ is a well defined real matrix. Taking this into account and that 
\be
g(s)^*=g(s)+i \frac{Q(s)}{4 \pi \sqrt{s}}
\ee
since 
\be
\hbox{Im} g(s)=-\frac{Q(s)}{8 \pi \sqrt{s}},
\ee
we can write eq. (\ref{uni3}) as:

\be
\label{uni4}
F(s)=\left[I+\widetilde{Q}(s)^{-1}\cdot K(s) \cdot \widetilde{Q}(s) \cdot g(s)\right]^{-1}\cdot
\left[ I+\widetilde{Q}(s)^{-1}\cdot K(s)\cdot \widetilde{Q}(s)\cdot g^*(s) \right]\cdot
F^*(s)  
\ee

Multiplying both sides of the former equation by $\left[I+\widetilde{Q}(s)^{-1}\cdot K(s) \cdot
\widetilde{Q}(s) \cdot g(s)\right]$ we arrive at the interesting result:
\be
\label{uni5}
\left[I+\widetilde{Q}(s)^{-1}\cdot K(s) \cdot \widetilde{Q}(s) \cdot g(s)\right] \cdot F(s)=
\left[I+\widetilde{Q}(s)^{-1}\cdot K(s) \cdot \widetilde{Q}(s) \cdot g^*(s)\right] \cdot F^*(s)
\ee 
This implies that the matrix $\left[I+\widetilde{Q}(s)^{-1}\cdot K(s) \cdot \widetilde{Q}(s) \cdot
g(s)\right] \cdot F(s)$ has no cuts at all, since the right hand cut or unitarity cut, the only one
present in $g(s)$ or $F(s)$, has been removed. Note that, as it is well known, a form
factor is an analytic function of $s$ except for the presence of the right hand cut from
threshold up to infinity. It can also have poles on the real $s$ axis below threshold,
corresponding to bound states, and poles in the complex plane in the unphysical Riemann
sheets corresponding to resonances. On the other hand, as discussed above in
contrast to $K(s)$, 
$\widetilde{Q}(s)^{-1}\cdot K(s) \cdot \widetilde{Q}(s)$ is a function free of physical
 or kinematical cuts. From eq. (\ref{uni5}) we write now

\ba
\label{uni6}
F(s)&=&\left[I+\widetilde{Q}(s)^{-1}\cdot K(s) \cdot \widetilde{Q}(s)\cdot g(s) 
\right]^{-1}\cdot R(s) \nonumber\\
&=&\widetilde{Q}(s)^{-1}\cdot \left[I+K(s)\cdot g(s)\right]^{-1} \cdot \widetilde{Q}(s) \cdot R(s)
\ea
with $R(s)$ a function free of any cut. Note also that the matrix $D(s)\equiv I+
K(s)\cdot g(s)$ was already introduced in eq. (\ref{T}) in relation with the purely strong
$T$-matrix.

As told above, the considered large $N_c$ leading contributions are given in eqs. 
(\ref{fftree}). In fact in this limit $D(s)=I+{\mathcal{O}}(N_c^{-1})$ and hence 
we have:

\be
\label{ggnc}
F(s)=R(s)_{N_c \,\,leading}=F^t(s)
\ee 
where $F^t(s)$ is the tree level form factor. As a result we can write:
\begin{equation}
\label{subalpha}
R(s)=F^{t}(s)+\delta R(s)
\end{equation}
with $\delta R(s)$ subleading in large $N_c$ and free of any cut. Since the pole
singularities present in the form factor correspond to tree level resonances 
(already present in $F^t(s)$) or dynamically generated resonances from the
$D(s)$ matrix, see eqs. (\ref{uni6}) and (\ref{T}), the $\delta R(s)$ matrix should 
then be simply a matrix made up of
polynomials\footnote{It should be understood in the former sentence that 
subleading corrections in $1/N_c$ to the couplings and masses of the resonances 
(not coming in our formalism by the $D(s)$ function), have to be reabsorbed in a 
renormalization of ${F}^{t}(s)$, eq. (\ref{fftree}).}.

        Hence, we can rewrite eq. (\ref{uni6}) as:

\begin{equation}
\label{finalff2}
F(s)=\left[I+\widetilde{Q}(s)^{-1}\cdot K(s) \cdot \widetilde{Q}(s)\cdot g(s)\right]^{-1} \cdot 
\left[ F^t(s)+\delta R(s) \right]
\end{equation}

We can further reduce eq. (\ref{finalff2}) for the two channel case. Indeed,
given the structure of the $K$ matrix of eq (\ref{treeT}) for $I=1$ and $\pi\pi$ and 
$K\bar{K}$ channels we find

\be
\label{matrix}
F(s)=\frac{1}{det[D(s)]}
\left( \begin{array}{cc}
1+p_{2}(s)^{2}\, \beta(s)\, g_{2}^{1}(s) & -\sqrt{2}p_{2}(s)^{2}\, \beta(s)\,
 g_{2}^{1}(s) \\
-\sqrt{2}p_{1}(s)^{2}\, \beta(s)\, g_{1}^{1}(s) & 1+2p_{1}(s)^{2}\, \beta(s)\, 
g_{1}^{1}(s)\\
\end{array} \right) \left(F^{t}(s)+\delta R(s) \right)
\ee
with $\beta$ given by
\be
\label{beta}
\beta(s)=\frac{1}{3f^{2}}\left(1+\frac{2G_{V}^{2}}{f^{2}}\frac{s}{M_{\rho}^{2}-s}\right)
\ee
and $det[D(s)]=1+2p_{1}(s)^{2}\, \beta(s)\, g^{1}_{1}(s)+p_{2}(s)^{2}\, \beta(s)\,
 g_{2}^{1}(s)$. 
For $I=0$, with only one channel, eq. (\ref{finalff2}) is simple enough. In the
explicit formula of eq. (\ref{matrix}) we can see that there are indeed no
kinematical cuts associated to the $p_{1}p_{2}$ factors of the matrix $K$, as
argued above.

The unitarity method we have introduced is similar to the one discussed 
in
ref. \cite{basdevant} in order to solve the Muskhelishvili-Omn\`es problem 
\cite{om}. 
The differences arise from the fact that we have showed how to deal with the kinematical
singularities associated with the product $p_i\, \hbox{Im}F(s)_i$ eq. (\ref{optiff}), 
and also because we have allowed the presence of pole 
singularities both in 
$D(s)$, due to $K(s)$, and in $\tilde{F}(s)$. However, they appear in such a way 
that they cancel each other and the results are finite. In that reference
\cite{basdevant} the connection
between eq. (\ref{finalff2}) and an Omn\`es representation for the
elastic case is also discussed. This relation is easily seen since in the elastic case 
$D^{-1}(s)$ has the phase of the strong scattering amplitude, eq. (\ref{T}), and only 
has the right hand cut. There is, however, a subtle point which is that $D^{-1}(s)$ has zeros for $s$ equal to the bare masses of the 
vector resonances introduced in $K(s)$ and they have to be removed in order to make
an Omn\`es representation. This is accomplished automatically in eq. (\ref{finalff2})
due to the explicit presence of the bare resonance poles in $F^t(s)$. 

\begin{figure}[ht]
\centerline{\includegraphics[width=0.7\textwidth]{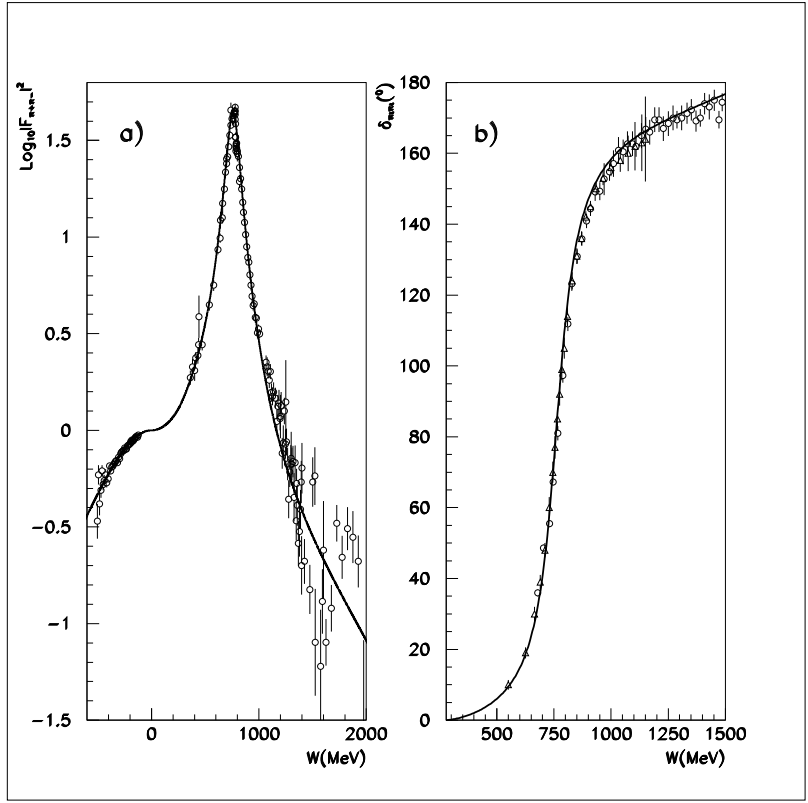}}
\caption{$W$ is defined as $\sqrt{s}$ for $s>0$ and as $-\sqrt{-s}$ for $s<0$. 
From left to right: a) $\pi^{+} \pi^{-}$ vector form factor. Experimental 
data from ref. \cite{barkov} collecting also data from refs. 
\cite{otros} for positive $s$. For negatives $s$ ref. \cite{nega} b) 
$\pi\pi$ P-wave phase shifts. Triangles from ref. 
\cite{estabrooks} and circles from ref. \cite{hyams}}
\label{tres}
\end{figure}

The degree of the polynomials present in $\delta R(s)$ can be restricted when 
considering the behavior of the form factors in the limit $s \rightarrow \infty$. 
In fact from eqs. (\ref{fftree}), (\ref{treeT}) and 
(\ref{finalff2}) we have the following limits:

\begin{eqnarray}
\label{limit}
\hbox{\Large{I=1}}&&\nonumber\\
F_1(\infty)&=&\frac{1-\frac{F_V G_V}{f^2}+\delta R^1_1(\infty)+g^1_2(\infty) 
\frac{s}{12f^2}(1-\frac{2 G_V^2}{f^2})\left[\delta R^1_1(\infty)-\sqrt{2}
\delta R^1_2(\infty)\right]}{1+\frac{s}{6f^2}(1-\frac{2
G_V^2}{f^2})\left\{g^1_1(\infty)+\frac{1}{2}g^1_2(\infty)\right\}} 
\nonumber \\
F_2(\infty)&=&\frac{1-\frac{F_V G_V}{f^2}+\sqrt{2}\delta R^1_2(\infty)+g^1_1(\infty) \frac{s}{6
f^2}(1-\frac{2 G_V^2}{f^2})\left[\sqrt{2} \delta R^1_2(\infty)-
\delta R^1_1(\infty)\right]}
{\sqrt{2}+\frac{\sqrt{2} s}{6f^2}(1-\frac{2
G_V^2}{f^2})\left\{g^1_1(\infty)+\frac{1}{2}g^1_2(\infty)\right\}} \nonumber \\
\hbox{\Large{I=0}}&&\nonumber\\
F_1(\infty)&=&\frac{1-\frac{F_V G_V}{f^2}+\sqrt{2}\,\delta R^0_1(\infty)}{\sqrt{2}+
\frac{\sqrt{2} s}{4 f^2}(1-\frac{2 G_V^2}{f^2})g^0_1(\infty)}
\end{eqnarray}
where $g^I_i(\infty)$ is a quantity logarithmically divergent with $s \rightarrow
\infty$, see eq. (\ref{g(s)}). We denote by $\delta R^I_i(\infty)$ the value of
the vector element of $\delta R(s)$ with isospin $I$ and channel $i$ for 
$s\rightarrow \infty$. Clearly, this value diverges as $s^{n(I;i)}$ with $n(I;i)$ the 
degree of the polynomial $\delta R^I_i(\infty)$.

From perturbative QCD \cite{rafael}, it is
known that the vector pion form factor goes at most as a constant for $s\rightarrow
\infty$. Although experimentally, and also from the quark counting rules, it is very likely 
that it vanishes for $s\rightarrow \infty$. Experimentally, we have $F_V=154$ MeV from the
observed decay rate $\Gamma(\rho^0 \rightarrow e^+e^-)$ \cite{ecker} and $G_V=53$ MeV 
from a study of the pion electromagnetic radii after taking into account corrections 
from chiral loops at next to leading order \cite{gl2}. With these values, we see from eqs.
(\ref{limit}) for the $I=0$ case, that $\delta R^0_1(s)$ can have, at most, degree one since 
otherwise the form factor would be divergent (we are assuming here that the high energy limit of
the kaon form factors is the same as the aforementioned one of the pion vector 
form factor). Analogously for 
$I=1$, we see that the difference $\delta R^1_1(s)-\sqrt{2}\delta R^1_2(s)$ can be at most a 
constant, while, independently, $\delta R^1_i(s)$ can have degree one. If we further require 
that the vector form factors vanish for $s\rightarrow \infty$, both for the pion and kaon ones, 
then 
$\delta R^I_i(s)$ are constants and 
\be
\label{deltas}
\delta R^1_1=\sqrt{2}\delta R^1_2.
\ee 
In the
following we will assume that this is the case\footnote{We refer the reader to
refs. 
\cite{dumm,ecker} which stress the importance to keep the short distance QCD constraints 
in order to guarantee the independence of the results under spin-1 field redefinitions.}. Note 
that since $\delta R^1_i(s)$ are at least ${\mathcal{O}}(p^2)$ quantities, these constants 
must be proportional to the square of the pseudoscalar masses $m_\pi^2$, 
$m_K^2$. Then we will finally have:

\begin{equation}
\label{finalff}
F(s)=\left[I+\widetilde{Q}(s)^{-1}\cdot K(s) \cdot \widetilde{Q}(s)\cdot g(s)\right]^{-1}\cdot 
\left(F^t(s)+\alpha_I \right)
\end{equation}
where $\alpha_I$ is just a constant for $I=0$ 
and for $I=1$, taking into account eq. (\ref{deltas}), $\alpha_1$ can be written as: 

\be
\label{alpha'}
\left(
\begin{array}{c}
\alpha'_1 \\
\alpha'_1/\sqrt{2}
\end{array}
\right)
\ee
with $\alpha'_1$ a constant.

%

\subsection{Matching with one loop $\chi PT$}

In order to fix the values of the constants $d^I_{i}$ of $g^I_i(s)$ and
$\alpha_{I}$ of eq. (\ref{finalff}) we match our results with those of $\chi PT$
up to one loop \cite{gl2}. The form factors from \cite{gl2}, together with eq. 
(\ref{isospin}) for the definition of the isospin states, are:

\begin{eqnarray}
\label{GLffs}
F_{\pi\pi}^{I=1}&=&1+2H_{\pi\pi}+H_{KK} \nonumber\\
F_{K\bar{K}}^{I=1}&=&\frac{1}{\sqrt{2}}\left(1+2H_{\pi\pi}+H_{KK}\right)\\
F_{K\bar{K}}^{I=0}&=&\frac{1}{\sqrt{2}}(1+3H_{KK})\nonumber
\end{eqnarray}
with the function $H_{ii}(s)$ given by:
\begin{eqnarray}
\label{H(s)}
H_{ii}(s)=\frac{2s}{3f^{2}}L^r_{9}(\mu)+\frac{s}{192\pi^{2}f^{2}}\left(2\sigma_{i}(s)^{2}-
\sigma_{i}^{3}(s)\log\frac{\sigma_{i}(s)+1}{\sigma_{i}(s)-1}-\frac{1}{3}+
\log\frac{\mu^{2}}{m_i^{2}}\right)
\end{eqnarray}

In order to make the matching with the one loop $\chi PT$ form factors, we will take 
into account the fact that for evaluating the contribution of the octet of vector resonances 
to the ${\mathcal{O}}(p^4)$ $\chi PT$ counterterms, for instance to $L^r_9(\mu)$, the masses of 
these resonances have to be taken in the chiral limit \cite{ecker}. We will denote this mass in
the following by $M_V\approx M_\rho$ \cite{ecker}. 
As noted in ref. \cite{ecker} the $L^r_9(\mu)$ is saturated by the meson vector resonance 
exchanges at a scale $\mu$ around the mass of the $\rho$, such that

\be
\label{ressat}
2 L_9(\mu\approx M^{physical}_\rho)=\frac{F_V G_V}{M_V^2}
\ee
we will also make use of this result in the following.

Let us consider first the $I=1$ case. Expanding our results, eq. (\ref{finalff}), up to one loop in 
the $\chi PT$ counting, we have:

\ba
\label{fpios}
F^{I=1}_{\pi\pi}(s)&=&1+\alpha'_1+\frac{F_V G_V s}{M_V^2 f^2}+\frac{s}{96 \pi^2 f^2}\left\{2 
\sigma_\pi(s)^2-\sigma_\pi(s)^2 d^1_1-\sigma_\pi(s)^3 \log \frac{\sigma_\pi(s)+1}{\sigma_\pi(s)-1}
\right\}+ \nonumber \\
&+&\frac{s}{192 \pi^2 f^2}\left\{2 \sigma_K(s)^2-\sigma_K(s)^2 d^1_2-\sigma_K(s)^3 \log 
\frac{\sigma_K(s)+1}{\sigma_K(s)-1} \right\}
\ea
while the one loop $\chi PT$ result is:
\ba
\label{GLffos}
F^{I=1}_{\pi\pi}(s)&=&1+\frac{2L^r_{9}(\mu)s}{f^{2}}+\frac{s}{96 \pi^2 f^2}\left\{2 
\sigma_\pi(s)^2-\sigma_\pi(s)^3 \log
\frac{\sigma_\pi(s)+1}{\sigma_\pi(s)-1}-\frac{1}{3}-\log\frac{m_\pi^2}{\mu^2}
\right\} +\nonumber \\
&+&\frac{s}{192 \pi^2 f^2}\left\{2 \sigma_K(s)^2-\sigma_K(s)^3 \log 
\frac{\sigma_K(s)+1}{\sigma_K(s)-1} -\frac{1}{3}-\log\frac{m_K^2}{\mu^2}\right\}
\ea

The matching for $K\bar{K}$ form factor in $I=1$ does not give any new condition
since this form factor, at the one loop chiral level, is simply $1/\sqrt{2}$ of
the $\pi\pi$ one, 
both in our approach and in $\chi PT$ \cite{gl2}. Note that this happens independently of the
value of the constant $\alpha'_1$, because of eq. (\ref{alpha'}).

In the following we will take 
\be
\label{alpha0}
\alpha'_1=0
\ee
in order to constraint further our approach. This can be
done since, as we show below, we can match our results with
one loop $\chi PT$ by choosing appropriate values for $d^1_1$ and $d^1_2$. On the other hand, as we 
will see in the next section, a very nice description of the pion form factor and of the
$\pi\pi$ P-wave phase shifts, see fig. \ref{tres},  is given without including this 
extra degree of freedom. In addition, we will discuss below the sensitivity of our results 
under changes of $\alpha'_1\neq 0$.

By matching eqs. (\ref{fpios}) and (\ref{GLffos}) with $\alpha'_1=0$, taking also into account eq. 
(\ref{ressat}), one finds the condition:

\be
\label{condition}
\sigma_\pi(s)^2 d^1_1+\frac{\sigma_K(s)^2}{2}d^1_2=\frac{1}{2}+\log\frac{m_\pi^2}{\mu^2}
+\frac{1}{2}\log\frac{m_K^2}{\mu^2}
\ee

Identifying the terms independent of $s$ and linear in $1/s$, we finally have:

\ba
\label{dpi}
d^1_1=\frac{m_K^2}{m_K^2-m_\pi^2}\left(\log\frac{m_\pi^2}{\mu^2}+\frac{1}{2}
\log\frac{m_K^2}{\mu^2}+\frac{1}{2}\right)\nn\\
d^1_2=\frac{-2\ m_\pi^2}{m_K^2-m_\pi^2}\left(\log\frac{m_\pi^2}{\mu^2}+\frac{1}{2}
\log\frac{m_K^2}{\mu^2}+\frac{1}{2}\right)
\ea

Next we do the matching in the $I=0$ sector. Our chiral one loop expression for this form factor is:

\be
\label{I0KK}
\sqrt{2}F^{I=0}_{K\bar{K}}=1+\sqrt{2} \alpha_0+\frac{F_V G_V s}{M_V^2f^2}+\frac{s}{64 \pi^2 f^2}\left\{
2 \sigma_K(s)^2-d^0_1 \sigma_K(s)^2-\sigma_K(s)^3 \log\frac{\sigma_K(s)+1}{\sigma_K(s)-1}\right\}
\ee
The $\chi PT$ one is:

\be
\label{I0KKchpt}
\sqrt{2}F^{I=0}_{K\bar{K}}=1+\frac{2 s}{f^2} L_9^r(\mu)+\frac{s}{64 \pi^2 f^2}\left\{2
\sigma_K(s)^2-\sigma_K(s)^3\log\frac{\sigma_K(s)+1}{\sigma_K(s)-1}-\frac{1}{3}-
\log\frac{m_K^2}{\mu^2} \right\}
\ee

By matching eqs. (\ref{I0KK}) and (\ref{I0KKchpt}), taking also into account eq. (\ref{ressat}), 
we obtain the condition:

\be
\label{condition0}
\sqrt{2} \alpha_0-\frac{s \sigma_K(s)^2 d^0_1}{64 \pi^2 f^2}=-\frac{s}{64 \pi^2 f^2}\left(\frac{1}{3}+
\log\frac{m_K^2}{\mu^2} \right)
\ee

Identifying the terms independent of $s$ and those linear in $s$:

\ba
\label{d_k0}
d^0_1=\frac{1}{3}+\log\frac{m_K^{2}}{\mu^2}\nn\\
\alpha_0=-\frac{m_K^{2}}{16\sqrt{2}\,\pi^{2}f^{2}}\left(\frac{1}{3}+\log\frac{m_K^{2}}{\mu^2}\right)
\ea

We see indeed that $\alpha_0$ is subleading in $N_{c}$ since $f^{2}\sim
N_{c}$ and $m_K^2 \sim 1$ in this counting. Numerically we find it to be of the order of 0.1 for 
$\mu=M_\rho$, consistent with the large $N_{c}$ counting.

\section{Results}

Once we have fixed the constants $d^I_i$ and $\alpha_I$ by eqs. (\ref{alpha0}), (\ref{dpi}) and 
(\ref{d_k0}), our final expression, eq. (\ref{finalff}), for 
the vector form factors only
depends on the bare masses of the $\rho$, $\phi$ and $\omega$ resonances and on
the couplings $G_V$ and $F_V$ and on $f$. The couplings $F_V$ and $G_V$ are fixed, as explained in
section 3, from their experimental values, 154 MeV and 53 MeV, respectively. The 
value for the parameter $f$ is taken from the second entry of ref. \cite{gl2}, where it is
derived by working out the relation of the ratio $f_\pi/f$ with the isoscalar
$\bar{u}u+\bar{d}d$ radius of 
the pion. The resulting central value is $f=87.4$ MeV. Notice that
this estimate is done for the value of $f$ in
the limit $m_u=m_d=0$ and $m_s\neq 0$ and there is some controversy about
the possible deviations with respect this value
when considering  the $SU(3)$ case \cite{Descotes:2000ct,Descotes:2000di}.
 The bare masses can be fixed in 
terms of $F_V$, $G_V$ and $f$ by the requirements
that the moduli of the $\pi\pi$ $I=1$ and $K\overline{K}$ $I$=0 P-wave amplitudes have a 
maximum for $\sqrt{s}=M^{physical}_\rho=770$ 
MeV and for $\sqrt{s}=M_{\phi}^{physical}=1019.413$ MeV, respectively. 
We obtain the values
$M_\rho=829.8$
MeV and $M_\phi=1026.581$ MeV, where the number of decimals correspond to the experimental
precision in which the physical masses are given \cite{pdg}.

For the mass of the $\omega$
we take directly 782 MeV since there are no experimental data in the region of
the $\omega$ and its contributions to other physical regions do not depend on
such fine details since the $\omega$ is very narrow.

Finally, note that our physical results does not depend on the 
regularization scale $\mu$ since any change in the scale is reabsorbed by the corresponding change in
the constants $d^I_i(\mu)$. However, since we have made use of eq. (\ref{ressat}), we have to 
consider a value for $\mu$ around the mass of the $\rho$ meson, where vector meson saturation of 
the $L^r_i(\mu)$ coefficients works \cite{ecker}. In the following we will fix $\mu=M_\rho$.

\begin{figure}[ht]
\centerline{\includegraphics[width=0.7\textwidth]{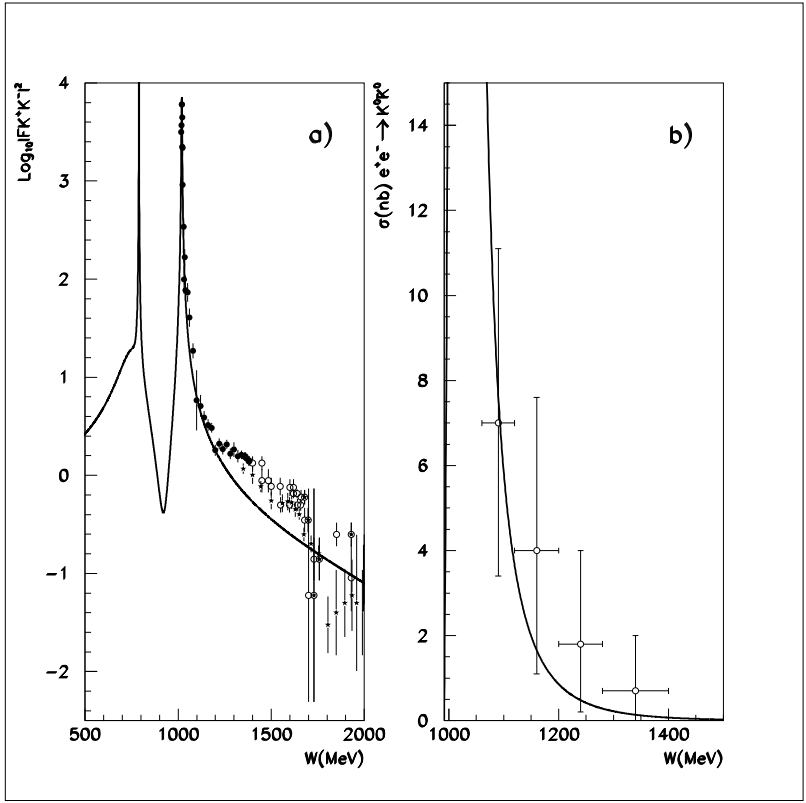}}
\caption{$W$ is defined as $\sqrt{s}$ for $s>0$ and as $-\sqrt{-s}$ for $s<0$. 
From left to right: a) $K^+K^-$ vector 
form factor. Experimental data: black circles \cite{ivanovkc}, white circles
\cite{dm1} and stars \cite{dm2}. b) $\sigma(e^+ e^- \rightarrow K^0\overline{K}^0)$ 
in nb. Data: \cite{ivanov82}}
\label{cuatro}
\end{figure}

\begin{figure}[ht]
\centerline{\includegraphics[width=0.7\textwidth]{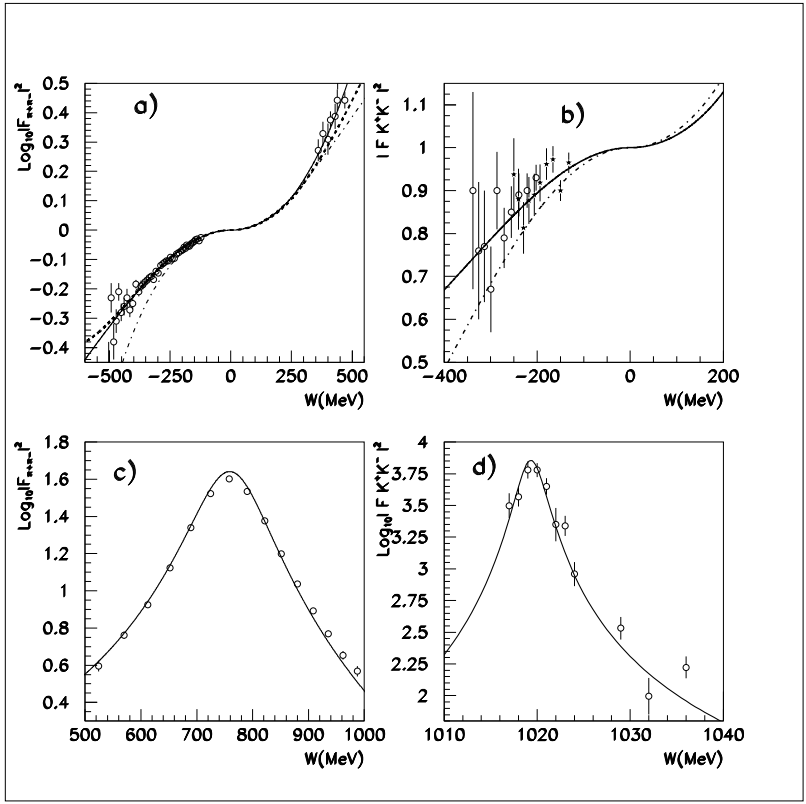}}
\caption{$W$ is defined as $\sqrt{s}$ for $s>0$ and as $-\sqrt{-s}$ for $s<0$. 
From left to right and top to bottom: a) Vector pion form factor. The dashed-dotted 
line represents one loop $\chi PT$ ref. \cite{gl2} and the dashed one the two loop $\chi PT$ 
result ref. \cite{cola}. b) $K^+ K^-$ form factor. The meaning of the lines is the same as before. Data are from 
ref. \cite{dali}, white circles, and the star points are from ref. \cite{amen}.
c)Vector pion form factor in the $\rho$ region without taking into account the
$\rho -\omega$ mixing effect. Data from $\tau$-decays
\cite{tau}. 
d) $K^+K^-$ form factor. Data \cite{ivanovkc}.}  
\label{cinco}
\end{figure}

        As can be seen in figs. \ref{tres}, \ref{cuatro} and \ref{cinco}, we can describe in a very precise
        way the vector pion 
and kaon form factors from negative values of $s$ up to about 1.44 GeV$^2$. 
The P-wave $\pi\pi$  phase shifts are also very well reproduced even for energies higher 
than those for the form factors, fig. \ref{tres} is displayed until $\sqrt{s}=1.5$ GeV. Note that the curves present in those 
figures do not result from a fit making use of our approach since the parameters
have been fixed in advance. Furthermore, the 
qualitative behavior of the data for the form factors is reproduced even for very high energies. 
For values of $\sqrt{s}$
higher than 1.2 GeV new effects appear: 1) the presence of more massive
resonances, $\rho'$, $\omega'$, $\phi'$... These states can be in principle
taken into account by our formalism just by adding more resonance to eqs. 
(\ref{fftree}) and (\ref{treeT}). 2) More seriously and less trivial is taking 
into account
the effect due to multiparticle states, e.g. $4\pi$, $\omega \pi$ ... In fact, since in 
our
model the widths of the resonances are derived dynamically in terms of the 
included channels we cannot mimic the effect of such multiparticle states as done in
other studies in which the widths of the resonances are fitted, that is,
incorporated by hand. This is currently done when using parameterizations of 
the type of ref. \cite{arcadi,barkov}. In fact, in the present case we have found that 
the effect of other channels cannot be neglected for energies higher than
$\sqrt{s}\gtrsim$ 1.2 GeV, for which we do
not give a fair reproduction of all the structures present in the data. We have
checked that the inclusion of higher mass resonances like the $\rho'$ improves
the agreement with the data but there are clear signs that other elements are
still missing.

In fig.\ref{cuatro}a one clearly sees the peaks of the $\phi$, $\rho$ and $\omega$ resonances, the
latter on the top of the $\rho$. The $\omega$ peak corresponds to that of a zero width
resonance since it appears below the $K\overline{K}$ threshold and the $3\pi$ state is
not included.

        In figs.\ref{cinco}a,b we compare for low energies our results for 
the pion and charged kaons vector form factors with the results obtained in $\chi PT$ 
for the pion vector form factor, up to one \cite{gl2} and two loops \cite{ulf2,cola}, and with 
the one loop $\chi PT$ \cite{gl2} charged kaon vector form factor. In the
figures the aforementioned matching, guaranteed by construction, is clearly 
seen, and the resummation accomplished by eq. (\ref{finalff})
provides the right corrections to the $\chi PT$ results also for low energies. In fact, we
see a much better agreement of our result with the two loop $\chi PT$ pion vector form
factor than with the one loop result. 

In table 1 we give the calculated
electromagnetic radii of the pions,
charged and neutral kaons and compare them with one loop 
$\chi PT$ \cite{gl2} and with experiment. We have not shown the two loop 
$\chi PT$  value for $<r_\pi^2>$ since its experimental value \cite{pira} is 
taken as an input in order to fix a counterterm \cite{ulf2}. In ref. \cite{cola} it is
argued that this is a sensible assumption within an accuracy around a $10 \%$ in the value 
of $<r^2>_V^\pi$. Our results for pions and charged kaons are compatible with the
$\chi PT$ results and experiment within errors. For the case of neutral kaons,
assuming as we have, the same uncertainties as in the $\chi PT$ results, we are
also compatible with experiment. It is also interesting to compare our values for the
electromagnetic radii with a low energy theorem due to Sirlin \cite{sirlin}. This
theorem states that,
\be
\label{sirlin}
{\mathcal{O}}[(m_s-\hat{m})^2]=\frac{1}{2}\left(F_V^\pi(s)+F_V^{K^+}(s)
\right)+F_V^{K^0}(s)-f_+^{K\pi}(s)
\ee
where $m_s$ is the mass of the strange quark, $\hat{m}$ is the average between the masses
of the up and down quarks and $f_+^{K\pi}$ is one piece of the vector $K\pi$ form
factor \cite{gl2}. In the previous relation the contribution of heavy quarks have been neglected and
we follow the notation of ref. \cite{gl2}. Numerically at ${\mathcal{O}}(p^4)$ in
$\chi PT$ \cite{gl2} one has, for the same combination of electromagnetic radii as in the
right hand side of eq. (\ref{sirlin}), $0.00\pm 0.05$ fm$^2$. At ${\mathcal{O}}(p^6)$ the result is
\cite{post} $0.021 \pm 0.003$ fm$^2$. From the values shown in table 1, taking from experiment
the electromagnetic radius $<r^2>_V^{K\pi}=0.36\pm 0.02$ fm$^2$ \cite{pdg}, as also done
in ${\mathcal{O}}(p^4)$ $\chi PT$, we have $-0.07 \pm 0.05$ fm$^2$. The experimental
value, from refs. \cite{amen,molzon,amen2,pdg}, is $-0.02 \pm 0.05$ fm$^2$. 

        In figs.\ref{cinco}c,d the resonance regions for the
$\rho$ and $\phi$ mesons are shown in detail. In the case of the pion form
factor in the $\rho$ region, we have compared our results with the experimental data of ref. \cite{tau} since in $\tau$ decays
only the $I=1$ part of the pion form factors plays a role and we are working in
the isospin limit. As can be seen the reproduction of data is very good in both
cases.

As announced before, we have also studied the dependence of the results on the subleading 
constant $\alpha'_1$ which has been set to zero in eq. (\ref{alpha0}). Notice that 
the matching to $\chi PT$ gives new values for the
$d^{1}_{I}$ parameters in terms of those of $\alpha'_1$ and also one has to recalculate the bare mass 
of the $\rho$ resonance. The new results, obtained with reasonable values of
$\alpha'_1$, can be hardly distinguished from the $\alpha'_1=0$ case in the
region $\sqrt{s}<1.2$ GeV. A fit to the $\pi \pi$ phase shifts and the form
factors data gives that the best values are $\alpha'_1=-0.0012$ and
$M_{\rho}=830.23$ MeV, leading to results that are indistinguishable from the
ones obtained by setting $\alpha'_1=0$.

        Finally, in our approach we have taken into account $\pi\pi$ and 
$K\overline{K}$ coupled channels for the $I=1$ P-wave. The effect of coupled
channels is more important for kaons, particularly for the neutral ones, than
for pions. This can be expected since when we decouple the kaons from the pions
the $\rho$ resonance appears in the kaon vector form
factors as a zero width resonance and this is a very bad approximation, particularly, 
if we are interested around the rather broad $\rho$ 
energy region. On the other hand, around the peak of the $\phi$ the effect of the pions is
small, of the order of a few percent, and the kaon vector form factor is dominated by the
$K\overline{K}$ elastic channel strongly coupled to the $\phi$. However, as we move 
again to higher energies, where the form factors
have dropped substantially, the pion channels become again important.

\begin{table}[ht]
\begin{center}
\begin{tabular}{|l|l|l|l|}
\hline
e.m. radii (fm$^2$) & Our results & One loop $\chi PT$ & Experiment \\ 
\hline
$<r^2>_V^{\pi}$ &  0.44  &  0.44 $\pm$ 0.04  & 0.439 $\pm$ 0.030 
\cite{amen2} \\
\hline
$<r^2>_V^{K^+}$ &  0.34  &  0.38 $\pm$ 0.03  & 0.34 $\pm$ 0.05 \cite{amen} \\
& & & 0.28 $\pm$ 0.07 \cite{dali} \nn \\ 
\hline
$<r^2>_V^{K^0}$ & -0.10  & -0.04 $\pm$ 0.03  & -0.054 $\pm$ 0.026 
\cite{molzon} \\
\hline
\end{tabular}
\caption{Electromagnetic radii for $\pi^+ \pi^-$, $K^+ K^-$ and $K^0
\overline{K}^0$, from left to right respectively. In the table are given our results,
the ones from $\chi PT$ at next-to-leading order ref. \cite{gl2} and some experimental data. 
The uncertainty in our values has to be considered of the same size than the one in $\chi
PT$, that is, around $\pm 0.04$ fm$^2$.}
\end{center}
\end{table}

\subsection{Isospin violation: $\rho -\omega$ mixing.}

The incorporation of isospin violation effects, due to the mass differences from
the $u$ and $d$ quarks and electromagnetic corrections, is readily possible, up
to some extend, within the present
formalism.  As an example, consider the OZI violating process $\phi \rightarrow \pi^{+}
\pi^{-}$ studied in ref. \cite{Oller:2000ag} within a chiral unitary approach. 
It was noted there that the $\omega
\rightarrow \pi^{+} \pi^{-}$ decay requires a direct $\rho -\omega$ mixing which
makes this process qualitatively different to the $\phi \rightarrow \pi^{+}
\pi^{-}$. This $\rho -\omega$ mixing has been studied in the context of chiral
Lagrangians, together with the large $N_c$ expansion, in \cite{Urech:1995ry} with the result:

\begin{center}
\begin{equation}
 i \mathcal{L} \rightarrow i \frac{1}{2}\tilde{\Theta}_{\rho\omega}\rho^0_{\mu\nu}
\textrm{ } \omega^{\mu\nu}
\label{mixing}
\end{equation}
\end{center}
with 

\begin{center}
\begin{equation}
\tilde{\Theta}_{\rho \omega}=-(m^{2}_{K^{0}}-m^{2}_{K^{+}})+(m^{2}_{\pi ^{0}}
-m^{2}_{\pi ^{+}})  +\frac{e^{2}F_{V}^{2}}{3}
\label{theta}
\end{equation}
\end{center}

\begin{figure}[ht]
\centerline{\includegraphics[width=0.3\textwidth]{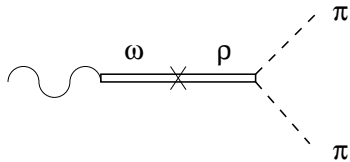}}
\caption{Feynman diagram corresponding to the $\rho -\omega$ mixing contribution
to the pion form factor.}  
\label{mixingdiag}
\end{figure}

Given the smallness of the contribution from $\rho -\omega$ mixing one can add
to the result for the pion form factor evaluated before the contribution from
the mechanism of fig.\ref{mixingdiag}.

The contribution of fig.\ref{mixingdiag} is straightforward and has been
evaluated previously in \cite{tesis}. With the $\gamma \omega$ and $\rho \pi
\pi$ couplings from \cite{ecker} and the $\rho -\omega$ coupling of
eq. (\ref{mixing}) we obtain:

\begin{center}
\begin{equation}
\Delta
F=-\frac{1}{3}\frac{F_{V}G_{V}}{f^{2}}s\frac{1}{s-M_{\omega}^{2}+i\sqrt{s}
\Gamma_{\omega}(s)}\textrm{ }\frac{1}{s-M_{\rho}^{2}+i\sqrt{s}
\Gamma_{\rho}(s)}\textrm{ }\tilde{\Theta}_{\rho \omega}
\end{equation}
\end{center}

\noindent Since this correction is only visible around the $\rho$ peak we take
for $\Gamma_{\omega}(s)$, $\Gamma_{\rho}(s)$ the constant values of the PDG
\cite{pdg} (using instead an energy dependence width of the $\rho$ meson does
not introduce any appreciable effects).
 
\begin{figure}[ht]
\centerline{\includegraphics[width=0.65\textwidth]{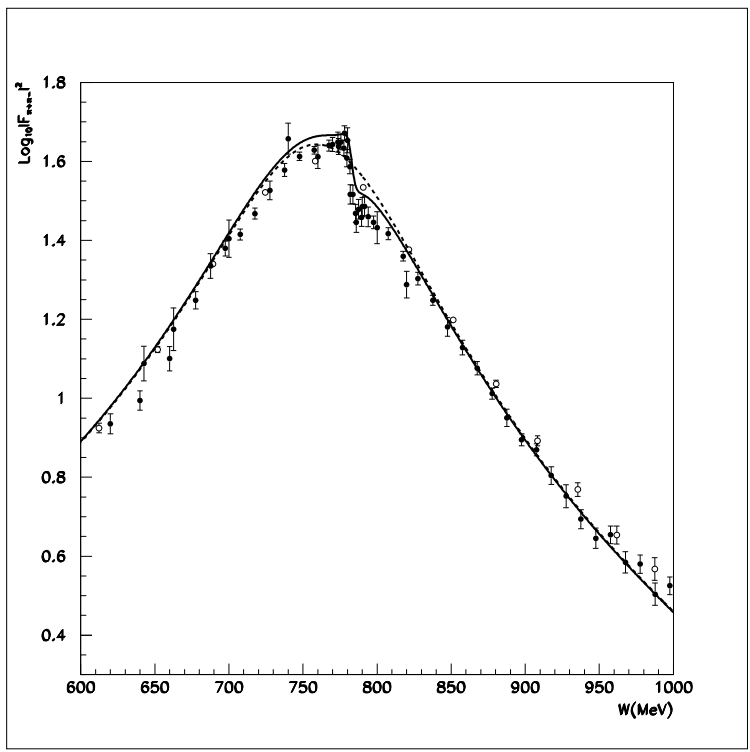}}
\caption{Solid line: pion vector form factor taking into account the $\rho
-\omega$ mixing; dashed line: I=1 vector form factor in the isospin limit. Data: black circles from
$e^{+} e^{-} \rightarrow \pi^{+} \pi^{-}$ reactions \cite{barkov,otros}; white
circles: data from $\tau$-decays
\cite{tau}. }  
\label{ffmixing}
\end{figure}

With a value for $\tilde{\Theta}_{\rho \omega}$ of $-4500$ MeV$^2$ from eq.
(\ref{theta}), which is also consistent with empirical determinations 
\cite{Gardner:1998ie,Williams:1998nj}, we get the form factor shown in fig.\ref{ffmixing} which
reproduces fairly well the experimental data at the peak of the pion form factor.
It is worth mentioning that this  effect of isospin violation 
 is not observed in the data from
$\tau$ decay since this process is only sensitive to the $I=1$ current (see
fig.\ref{cinco}). 

\section{Conclusions}

We have developed an appropriate method to take into account the final state
interaction corrections to the tree level amplitudes of refs. \cite{gl2,ecker} with
explicit resonance fields. 

The strong tree level scattering amplitudes from refs. \cite{gl2,ecker} were unitarized 
in coupled channels following the 
scheme developed in ref. \cite{nd}. These amplitudes were then implemented in
order to derive the final state corrections to the tree level vector form
factors, eqs. (\ref{fftree}). As a result a very good description
of data is accomplished for $\sqrt{s}\lesssim$ 1.2 GeV. For higher energies multiparticle
channels are no longer negligible and also new resonances appear, although our results 
still give the trend of the experimental data.

Our expressions reproduce the $\chi PT$ expansion of the pion and kaon vector form
factors up to one loop \cite{gl2}. The resummation of our scheme leads, in the low energy 
region where $\chi PT$ is expected to hold (see fig.\ref{cinco}a), to a much better 
agreement with the two loop $\chi PT$ pion vector form factor than with the one loop one. 
 However, at the 
 same time, we are able to provide a very good description at higher energies, including the 
 resonance regions where standard $\chi PT$ could not be applied, figs.
 \ref{cinco}c and
 \ref{cinco}d. Furthermore, our calculated electromagnetic
 radius of the $\pi^+$, $K^+$ and $K^0$, table 1, are in agreement with
 experiment, assuming the same uncertainties as in the $\chi PT$ results. 
 
 We have also taken into account the $\rho -\omega$ mixing as a main source of
 isospin breaking in order to compare our results with the form factor obtained
 from $e^{+}e^{-} \rightarrow \pi^{+} \pi^{-}$. The input for the $\rho -\omega$
  mixing has been taken from recent work respecting
 also chiral symmetry in the same
line as the rest of the Lagrangians employed here.
 
 Finally, our method also
 provides P-wave scattering amplitudes for pions and kaons in very good agreement with
 the experimental data on the phase shifts for the former one. For the case of the kaons
 there are no scattering data
 available, but they would be highly desirable as a further test of the chiral
 unitary approach followed here.

\bigskip

\subsection*{Acknowledgments}
We would like to acknowledge useful discussions and a critical reading by 
Ulf-G. Mei{\ss}ner. Multiple and useful discussions with A. Pich are also 
acknowledged. We also thank J. Portol\'es for calling our attention to ref. 
\cite{dumm}. Partial financial support from the DGICYT under
 contract PB96-0753 and from the EU TMR network Eurodaphne, contract no. 
ERBFMRX-CT98-0169 is acknowledged as well. J.E.P. thanks financial support 
from the Ministerio de Educaci\'on y Cultura.

\end{document}